\documentstyle[11pt, titlepage]{article}

\newcommand{\beq}{\begin{equation}}
\newcommand{\beqn}{\begin{eqnarray}}
\newcommand{\eeq}{\end{equation}}
\newcommand{\eeqn}{\end{eqnarray}}
\begin{document}
\begin{titlepage}
\rightline {HUTP-95/A024}
\rightline {DART-HEP-95/02}
\rightline {astro-ph/9507048}
\rightline {Submitted to {\it Phys. Lett. B}}
\begin{center}
\bigskip \bigskip \bigskip \bigskip \bigskip
\Large\bf Frame-Independent Calculation of \\
Spectral Indices from Inflation \\
\bigskip\bigskip\bigskip
\normalsize\rm
David I. Kaiser \\
\bigskip \it Lyman Laboratory of Physics\\
Harvard University\\
Cambridge, MA 02138\\ \rm
e-mail:  dkaiser@fas.harvard.edu\\
\bigskip\bigskip\bigskip\bigskip
12 July 1995\\ \bigskip
\bigskip\bigskip\bigskip\bigskip
\end{center} \narrower
\bf Abstract \rm  Spectral indices from models of inflation which
incorporate a Generalized Einstein Theory (GET) gravity sector are
calculated to first order in a slow-roll expansion.  By quantizing a
suitably-generalized measure of the intrinsic curvature perturbation, the
spectral indices as calculated in the Jordan frame now match those as
calculated following a conformal transformation, in the Einstein frame.
 \bigskip\\

\end{titlepage}
\newpage

\baselineskip 24pt
\section{Introduction}
\indent  Over the past few years, a formalism has been developed for
calculating the spectral index ($n_s$) of the primordial density
perturbation, based on an expansion in
inflationary \lq\lq slow-roll"
parameters.~\cite{SL}~\cite{KV}~\cite{psra}~\cite{LidLyth}  This
formalism assumes that the gravitational portion of the action takes the
canonical Einstein-Hilbert form.  As demonstrated in~\cite{ngets}, this
formalism may be applied to inflation models which incorporate a
Generalized Einstein Theory (GET) gravitational action if use is made of
a conformal transformation, which puts the action into the form of an
Einstein-Hilbert gravitational sector with a minimally-coupled scalar
field.  However, the gauge-invariant
measure of the intrinsic curvature perturbation upon which this
formalism is based is {\it not invariant} with respect to such a conformal
transformation.  As discussed in section 3.2 of~\cite{ngets},
therefore, under
certain initial conditions the spectral index as calculated in the Jordan
frame differs from the spectral index as calculated following the
conformal transformation, in the Einstein frame.  In this Letter, we
present a new means of calculating $n_s$ for these GET models of
inflation, the results of which are {\it both} gauge-invariant and
frame-independent.  By exploiting this new slow-roll expansion, the
discrepancy in
the spectral index between the two frames may be eliminated.  \\
\indent  Throughout this Letter, we will assume that the background
spacetime is that of a flat
Friedmann-Robertson-Walker line element:
\beq
ds^2 = g_{\mu\nu} dx^{\mu} dx^{\nu} = - dt^2 + a^2 (t) d\vec{x}^2 .
\label{ds}
\eeq
The Hubble parameter is then $H \equiv \dot{a} / a$,
where overdots denote derivatives with respect to $t$. \\
\indent  In section 2, we derive slow-roll expansions with which to
calculate $n_s$ for four distinct models of inflation.  First to be
treated is a model with an Einstein-Hilbert gravitational action and a
minimally-coupled scalar field.  Next, three closely-related GET models
of inflation are considered:  Induced-gravity Inflation (IgI), a theory
with a nonminimally-coupled scalar field (NMSF), and a general
scalar-tensor (GST) theory.  In section 3, we evaluate $n_s$ in both the
Jordan and Einstein frames for the case of IgI, and show that the
spectral indices now agree in the two frames.  Concluding remarks follow
in section 4.

\section{Calculating the spectral index}
\indent  The calculation of $n_s$ here follows closely the derivation by
Stewart and Lyth for the case of a canonical gravity sector and a
minimally-coupled scalar field~\cite{SL}.  Their treatment can be
extended to GET models of inflation by quantizing a suitably-generalized
measure of the intrinsic curvature perturbation.  Here we turn to the
gauge-invariant potential introduced by Bardeen~\cite{Bardeen}, $\Phi_H$,
which can be related to the intrinsic curvature perturbation for GETs
(see~\cite{Bardeen}~\cite{Hcqg}~\cite{Hapj}; henceforth we will drop the
subscript \lq\lq {\it H}").
Hwang has derived the field equations which $\Phi$ obeys in many specific
GETs~\cite{Hcqg}, and has further introduced variables~\cite{Hprd} with
which to cast these field equations into the same form as those studied
in~\cite{SL}.  By defining new slow-roll parameters based on these
variables, $n_s$ may be calculated unambiguously for GET models of
inflation. \\
\indent  First we consider the same model as that treated in~\cite{SL},
Einstein gravity with a minimally-coupled scalar field ($\phi$), but by
means of the potential $\Phi$.  In analogy with~\cite{SL}~\cite{LidLyth},
we may define
\beqn
\nonumber \Phi &\equiv& \int \frac{d^3 \vec{k}}{(2\pi)^{3/2}} \>
\Phi_{\vec{k}} (\eta) \> e^{i \vec{k} \cdot \vec{x}} , \\
\langle \Phi_{\vec{k}} \> \Phi_{\vec{\ell}}^* \rangle &\equiv&
\frac{2\pi^2}{k^3} \> P_{\Phi} \> \delta^3 (\vec{k} - \vec{\ell}) ,
\label{Phi}
\eeqn
where $\eta$ is the conformal time, defined as $d\eta \equiv a^{-1} dt$.
Introducing the variables $u$ and $z$ as~\cite{Hprd}
\beq
u \equiv \frac{\Phi}{\dot{\phi}} \>\> , \>\> z \equiv \frac{H}{a
\dot{\phi}} ,
\label{uvMSF}
\eeq
we may construct a quantum operator $\hat{u}(x)$ as
\beq
\hat{u} (x) = \int \frac{d^3 \vec{k}}{(2 \pi)^{3/2}} \left[ u_k (\eta)
\> \hat{a}_{\vec{k}} \> e^{i \vec{k} \cdot \vec{x}} + u_k^* (\eta)
\> \hat{a}^{\dagger}_{\vec{k}}\> e^{-i \vec{k} \cdot \vec{x}} \right] ,
\label{hatu}
\eeq
where the creation and annihilation operators obey the usual commutation
relations:
\beqn
\nonumber \left[ \hat{a}_{\vec{k}} \> , \> \hat{a}_{\vec{\ell}} \right]
&=& \left[ \hat{a}^{\dagger}_{\vec{k}} \> , \>
\hat{a}^{\dagger}_{\vec{\ell}} \right] = 0 , \\
\left[ \hat{a}_{\vec{k}} \> , \> \hat{a}^{\dagger}_{\vec{\ell}} \right]
&=& \delta^3 ( \vec{k} - \vec{\ell} ) \> , \>\> {\rm with}
\>\> \hat{a}_{\vec{k}} \vert 0 > = 0 .
\label{comrels}
\eeqn
Denoting $d/d\eta$ by a prime, the mode functions $u_k$ then
obey~\cite{Hprd}:
\beq
u^{\prime\prime}_k + \left( k^2 - \frac{z^{\prime\prime}}{z} \right) u_k
 = 0
\label{upp}
\eeq
in a flat Friedmann universe.  This equation is of exactly the same form
as that presented by Mukhanov~\cite{Muk} for studying
cosmological perturbations,
and upon which the derivation by Stewart and Lyth~\cite{SL} is based.
Note, however, that the variable $z$ in equation (\ref{uvMSF}) is the
inverse of Mukhanov's corresponding variable, $a \dot{\phi} / H$. \\
\indent  In order to solve equation (\ref{upp}) for $u_k$, we follow Stewart
and Lyth~\cite{SL} by introducing the slow-roll
parameters\footnote{Note that the slow-roll parameters $\epsilon$ and
$\delta$ as defined in equation (\ref{epsdel}) are often rewritten,
using the equations of motion for $H$ and $\phi$, in
terms of either $d H / d\phi$ and
higher derivatives (this is the so-called \lq\lq Hubble slow-roll
approximation" of~\cite{psra}), or in terms of $dV / d\phi$ and higher
derivatives (the \lq\lq Potential slow-roll approximation"~\cite{psra}).}
\beq
\epsilon \equiv - \frac{\dot{H}}{H^2} \>\> , \>\> \delta \equiv
\frac{\ddot{\phi}}{H \dot{\phi}} .
\label{epsdel}
\eeq
Then, from the definition of $z$ in equation (\ref{uvMSF}), the
$z^{\prime\prime}/ z$ term in equation (\ref{upp}) may be written
\beq
\frac{z^{\prime\prime}}{z} = a^2 H^2 \left[ \left( 2\epsilon + \delta
\right) \left(1 + \epsilon + \delta \right) - \frac{1}{H} \left(
\dot{\epsilon} + \dot{\delta} \right) \right] .
\label{zpp}
\eeq
During inflation, $\vert \epsilon \vert \> , \> \vert \delta \vert \ll 1$.
Furthermore, both
$\dot{\epsilon}$ and $\dot{\delta}$ are second-order in $\epsilon$ and
$\delta$; to first order, then, it is consistent to set $\dot{\epsilon} =
\dot{\delta} = 0$, which we will do here.  The conformal time $\eta$
thus takes the closed-form expression:
\beq
\eta = - \frac{1}{aH} \frac{1}{1 - \epsilon} ,
\label{eta}
\eeq
so that equation (\ref{zpp}) may be rewritten as:
\beq
\frac{z^{\prime\prime}}{z} = \frac{1}{\eta^2} \left( \nu^2 - \frac{1}{4}
\right) ,
\label{zpp2}
\eeq
with
\beq
\nu = \frac{1}{2} + \frac{ 2\epsilon + \delta}{1 - \epsilon} .
\label{nuMSF}
\eeq
Equation (\ref{upp}) is now simply Bessel's equation; the mode functions
$u_k$ may be written in terms of Hankel functions as
\beq
u_k (\eta) = \left( - \eta \right)^{1/2} \left[ A_k H_{\nu}^{(1)} (- k \eta)
+ B_k H_{\nu}^{(2)} (- k \eta) \right] .
\label{u1}
\eeq
Requiring that $\hat{u}(x)$ behave as a free quantum field for $k / (aH)
\gg 1$ ({\it i.e.}, $u_k \rightarrow (2k)^{-1/2} e^{-ik\eta}$) sets:
\beq
B_k = 0 \>\> , \>\> A_k = \frac{\sqrt{\pi}}{2} \exp \left[ i \frac{\pi}{2}
\left( \nu + \frac{1}{2} \right) \right] .
\label{AB}
\eeq
In the opposite, long-wavelength limit ($k/(aH) \ll 1$), the mode
functions $u_k$ thus behave as
\beq
u_k \propto k^{ - \nu} .
\label{u2}
\eeq
{}From equations (\ref{Phi}), (\ref{uvMSF}), and (\ref{hatu}), the
vacuum expectation value for the quantum operator $\hat{\Phi}$ behaves as
\beq
\langle 0 \vert \hat{\Phi}_{\vec{k}}\> \hat{\Phi}_{\vec{\ell}} \vert 0
\rangle \propto \vert u_k \vert^2 \delta^3 ( \vec{k} -
\vec{\ell} ) ,
\label{PhiPhi}
\eeq
giving
\beq
P_{\Phi}^{1/2} (k) \propto k^{3/2 - \nu} .
\label{PPhi1}
\eeq
The large-scale behavior of $\Phi$ is considered
in~\cite{Hcqg}~\cite{Hprd}~\cite{Hapj}.  Setting ${\cal R}_{\vec{k}}
\propto k^{-1} \Phi_{\vec{k}}$ at this scale, where
${\cal R}$ is the intrinsic curvature perturbation, gives $P_{\cal
R}^{1/2} \propto k^{-1} P_{\Phi}^{1/2} \propto k^{1/2 - \nu}$.
The spectral index is defined by~\cite{SL}~\cite{LidLyth}
\beq
n_s \equiv 1 + \frac{d \ln P_{\cal R}}{d \ln k} ,
\label{ns}
\eeq
so, from equation (\ref{PPhi1}),
\beq
n_s = 2 - 2\nu .
\label{ns2}
\eeq
To first order, $\nu \simeq 1/2 + 2 \epsilon + \delta$, so that equation
(\ref{ns2}) may be rewritten
\beq
n_s \simeq 1 - 4 \epsilon - 2 \delta .
\label{firstO}
\eeq
This is the standard first-order result
for Einstein gravity with a minimally-coupled scalar
field.~\cite{SL}~\cite{psra}~\cite{LidLyth}  Given $\vert \epsilon \vert \>
, \> \vert \delta \vert \ll 1$ during inflation, it is clear that
such inflationary models generically predict density perturbation spectra
which are close to the $n_s = 1.00$ scale-invariant (Harrison-Zel'dovich)
spectrum. \\
\indent  The foregoing derivation may now be repeated for the three GET
models of inflation.\footnote{Note that all three of the GET models
considered here involve only a single scalar field, thereby avoiding the
nonadiabatic \lq\lq frictional damping" which arises in GETs which employ
more
than one scalar field.~\cite{GBW}  For more on the calculation of spectral
indices
from inflationary models with several dynamical degrees of freedom,
see~\cite{SasS}.  Metric perturbations in string cosmologies with $d$
spatial dimensions and $n$ internal dimensions are considered
in~\cite{BGGMV}. }  First we consider Induced-gravity
Inflation (IgI), the action for which may be written~\cite{IgI}
\beqn
\nonumber  S &=& \int d^4 x \sqrt{-g} \left[ \frac{1}{2} \xi \phi^2 R -
\frac{1}{2} \phi_{; \> \mu} \phi^{; \> \mu} - V (\phi) \right], \\
V(\phi) &=& \frac{\lambda}{4} \left( \phi^2 - v^2 \right)^2 ,
\label{Sigi}
\eeqn
where $\xi$ ($> 0$) is the nonminimal coupling strength, and is related
to the Brans-Dicke~\cite{BD} parameter $\omega$ by $\xi = (4
\omega)^{-1}$.  For IgI it is convenient to describe the equation of
motion for $\Phi$ in terms of the generalized variables $u$ and
$z$~\cite{Hprd}
\beq
u \equiv \frac{\phi^2}{\dot{\phi}} \> \tilde{\Phi} \>\> , \>\> z \equiv
\frac{H}{a \dot{\phi}} \left( 1 + \frac{\dot{\phi}}{H \phi} \right) ,
\label{uvIgI}
\eeq
where $\tilde{\Phi}$ is the potential $\Phi$ after a conformal
transformation has been performed, which puts the action of equation
(\ref{Sigi}) into the form of Einstein gravity with a (newly-defined)
minimally-coupled scalar field; this conformal transformation will be
considered below, in section 3.  From equation (\ref{uvIgI}) we may
define $\hat{u}(x)$ exactly as in equations (\ref{hatu}) and
(\ref{comrels}).  The new mode functions $u_k$ then obey the same
equation of motion as in equation (\ref{upp}), with $z$ now given by
equation (\ref{uvIgI}).~\cite{Hprd} \\
\indent  In order to solve for $u_k$, we
again define the two slow-roll parameters $\epsilon$ and $\delta$ as in
equation (\ref{epsdel}), and define a third slow-roll parameter:
\beq
\alpha \equiv \frac{\dot{\phi}}{H \phi} .
\label{alpha}
\eeq
In terms of $\epsilon$, $\delta$, and $\alpha$, the $z^{\prime\prime} /
z$ term in equation (\ref{upp}) may be written:
\beq
\frac{z^{\prime\prime}}{z} = a^2 H^2 \left[ \left( 2 \epsilon + \delta
\right) \left(1 + \epsilon + \delta \right) - \frac{1}{H} \left(
\dot{\epsilon} + \dot{\delta} \right) \right] - a^2 H
\frac{\dot{\alpha}}{(1 + \alpha)} \left[ 1 + 2 \epsilon + 2 \delta -
\frac{\ddot{\alpha}}{H \dot{\alpha}} \right] .
\label{zppIgI}
\eeq
As demonstrated in~\cite{ngets}, $\vert \alpha \vert \ll 1$ during
inflation.  Also,
as for $\dot{\epsilon}$ and $\dot{\delta}$, $\dot{\alpha}$ is
second-order in the slow-roll parameters:
\beq
\frac{\dot{\alpha}}{H} = \alpha \left( \delta + \epsilon - \alpha \right) ,
\label{dotalpha}
\eeq
so that to first order we may assume $\dot{\epsilon} = \dot{\delta} =
\dot{\alpha} = 0$.  From equation (\ref{dotalpha}), this gives $\alpha =
\epsilon + \delta$ to first order for IgI.  Taking $\dot{\epsilon} = 0$
means that the conformal time $\eta$ is again given by equation
(\ref{eta}), so that equation (\ref{zppIgI}) again reduces to
equation (\ref{zpp2}), with
\beq
\nu = \frac{1}{2} + \frac{\epsilon + \alpha}{1 - \epsilon} .
\label{nuIgI}
\eeq
Proceeding as above in equations (\ref{u1}) to (\ref{ns}) we again arrive
at $n_s = 2 - 2 \nu$, with $\nu$ now given by equation (\ref{nuIgI}).
Approximating $\nu \simeq 1/2 + \epsilon + \alpha$ to first order gives
\beq
n_s \simeq 1 - 2 \epsilon - 2 \alpha
\label{nsIgI}
\eeq
for IgI.  As for the case of a minimally-coupled scalar field with an
Einstein-Hilbert gravitational action, the spectral index for IgI thus
remains close to the $n_s = 1.00$ scale-invariant spectrum.  Before
evaluating $n_s$ for IgI (in section 3), we will next derive spectral
indices for two other common GET models of inflation. \\
\indent  The next GET model of inflation to be considered is that of a
nonminimally-coupled scalar field (NMSF), the action for which may be
written (see, {\it e.g.},~\cite{NMSF})
\beq
S = \int d^4 x \sqrt{-g} \left[ \left( \frac{1 + \kappa^2 \xi \phi^2}{2
\kappa^2} \right) R - \frac{1}{2} \phi_{; \> \mu} \phi^{; \> \mu} - V
(\phi) \right] ,
\label{Snmsf}
\eeq
where $V(\phi)$ can take a simple polynomial form, such as $V = \lambda
\phi^4$, or can be of the Ginzburg-Landau form, as in equation
(\ref{Sigi}).  Here $\kappa^2 \equiv 8 \pi G = 8 \pi M_{pl}^{-2}$, where
$M_{pl}
\simeq 1.22 \times 10^{19}$ GeV is the present value of the Planck mass.
The sign of $\xi$ in equation (\ref{Snmsf}) is chosen to match that
in~\cite{ngets}, and is the opposite of Hwang's choice in~\cite{Hprd}.
Following Hwang, we define the two quantities $F$ and $E$ as (noting the
new sign of $\xi$):
\beq
F \equiv 1 + \xi \phi^2 \>\> , \>\> E \equiv 1 + \xi \phi^2 (1 + 6 \xi) ,
\label{FE}
\eeq
in terms of which the appropriate $u$ and $z$ variables may be
written~\cite{Hprd}
\beq
u \equiv \frac{1}{\dot{\phi}} \sqrt{\frac{F^3}{E}} \> \tilde{\Phi} \>\> ,
\>\> z \equiv \frac{H}{a \dot{\phi}} \sqrt{\frac{F}{E}} \left( 1 +
\frac{1}{2}\frac{\dot{F}}{HF} \right) .
\label{uvNMSF}
\eeq
With these definitions for $u$ and $z$, we again construct $\hat{u}(x)$
as in equations (\ref{hatu}) and (\ref{comrels}), with the mode functions
$u_k$ obeying equation (\ref{upp}).  Alongside $\epsilon$ and $\delta$ we
now add the two slow-roll parameters
\beq
\beta \equiv \frac{1}{2}\frac{\dot{F}}{HF} \>\> , \>\> \gamma \equiv
\frac{1}{2}\frac{\dot{E}}{HE} ,
\label{betgam}
\eeq
with which the $z^{\prime\prime} / z$ term in equation (\ref{upp}) may be
written
\beqn
\nonumber  \frac{z^{\prime\prime}}{z} &=& a^2 H^2 \left[ \left(2 \epsilon +
\delta + \gamma - \beta  \right) \left( 1 + \epsilon + \delta +
\gamma - \beta  \right) - \frac{1}{H} \left( \dot{\epsilon}
+ \dot{\delta} +  \dot{\gamma} - \dot{\beta} \right) \right] \\
&-& a^2 H \frac{\dot{\beta}}{( 1 + \beta ) } \left[ 1 + 2 \epsilon + 2
\delta + 2 \gamma - 2 \beta - \frac{\ddot{\beta}}{H \dot{\beta}} \right] .
\label{zppNMSF}
\eeqn
Once again setting $\dot{\epsilon} = \dot{\delta} = \dot{\beta} =
\dot{\gamma} = 0$ to first order in these four slow-roll parameters,
equation (\ref{zppNMSF}) reduces to equation (\ref{zpp2}), with
\beq
\nu = \frac{1}{2} + \frac{2\epsilon + \delta +  \gamma - \beta }{1
- \epsilon} ,
\label{nuNMSF}
\eeq
giving, to first order,
\beq
n_s \simeq 1 - 4 \epsilon - 2 \delta - 2 \gamma + 2 \beta
\label{nsNMSF}
\eeq
for the spectral index from an inflationary model with a
nonminimally-coupled scalar field. \\
\indent  Lastly, we consider a general scalar-tensor theory (GST), the
action for which is a generalization of the original Brans-Dicke
theory~\cite{BD}:
\beq
S = \frac{1}{16\pi} \int d^4 x \sqrt{-g} \left[ \phi R - \frac{ \omega
(\phi)}{\phi} \phi_{; \> \mu} \phi^{; \> \mu} - V (\phi) \right] .
\label{Sgst}
\eeq
Again following Hwang~\cite{Hprd}, we define $w \equiv \omega + 3/2$, and
write $u$ and $z$ as
\beq
u \equiv \frac{1}{\dot{\phi}} \sqrt{\frac{\phi^3}{w}} \> \tilde{\Phi}
\>\> , \>\> z \equiv \frac{H}{a \dot{\phi}} \sqrt{\frac{\phi}{w}} \left( 1
+ \frac{1}{2}\frac{\dot{\phi}}{H \phi} \right) ,
\label{uvGST}
\eeq
so that the mode functions $u_k$ of $\hat{u}(x)$ obey equation
(\ref{upp}), with $z$ now given by equation (\ref{uvGST}).  To
$\epsilon$ and $\delta$, as defined in equation (\ref{epsdel}), and
$\alpha$, as defined in equation (\ref{alpha}), we now add the slow-roll
parameter $\zeta$
\beq
\zeta \equiv \frac{\dot{w}}{Hw} ,
\label{w}
\eeq
with which the $z^{\prime\prime} / z$ term in equation (\ref{upp}) may be
written
\beqn
\nonumber \frac{z^{\prime\prime}}{z} &=& a^2 H^2 \left[ \left( 2 \epsilon +
\delta +
\frac{1}{2} ( \zeta - \alpha ) \right) \left( 1 + \epsilon + \delta +
\frac{1}{2} ( \zeta - \alpha ) \right) - \frac{1}{H} \left(
\dot{\epsilon} + \dot{\delta} + \frac{1}{2} \left( \dot{\zeta} -
\dot{\alpha} \right) \right) \right] \\
&-& a^2 H \frac{\dot{\alpha}}{2 (1 + \alpha / 2 )} \left[ 1 + 2 \epsilon + 2
\delta + \zeta - \alpha - \frac{\ddot{\alpha}}{H\dot{\alpha}} \right] .
\label{zppGST}
\eeqn
Taking $\dot{\epsilon} = \dot{\delta} = \dot{\alpha} = \dot{\zeta} = 0$
to first order, equation (\ref{zppGST}) reduces to equation (\ref{zpp2}),
with
\beq
\nu = \frac{1}{2} + \frac{2\epsilon + \delta + (\zeta - \alpha ) / 2}{1 -
\epsilon} .
\label{nuGST}
\eeq
To first order in the four slow-roll parameters, this yields the spectral
index for a general scalar-tensor theory:
\beq
n_s \simeq 1 - 4\epsilon - 2 \delta - \zeta + \alpha ,
\label{nsGST}
\eeq
again close to the $n_s = 1.00$ scale-invariant spectrum. \\
\indent  The results for the spectral index from Induced-gravity
Inflation (equation \ref{nsIgI}), from a theory with a nonminimally-coupled
scalar field (equation \ref{nsNMSF}), and from a general scalar-tensor
theory (equation \ref{nsGST}) all apply to the Jordan frame for these
GETs; that is, for when the models are specified with their explicit
nonminimal $\phi R$ coupling, as in equations (\ref{Sigi}),
(\ref{Snmsf}), and (\ref{Sgst}).  In the next section, we demonstrate
for the case of Induced-gravity Inflation that this Jordan-frame
formalism for $n_s$ yields the same results for the spectral index as
those obtained in the Einstein frame, after use has been made of a
conformal transformation.

\section{Spectral Index from IgI in Jordan and Einstein frames}
\indent  We now compare the calculation of $n_s$ in the Jordan and
Einstein frames of IgI;  it was in the context of this model that the
discrepancy was first discussed, in section 3.2 of~\cite{ngets}.  The
action in equation (\ref{Sigi}) yields the coupled field
equations in the Jordan frame (for a flat Friedmann universe):
\beqn
\nonumber H^2 &=& \frac{1}{3\xi\phi^2} \> V(\phi) + \frac{1}{6\xi} \left(
\frac{ \dot{\phi}}{\phi} \right)^2 - 2H \left( \frac{\dot{\phi}}{\phi}
\right) , \\
\ddot{\phi} + 3H \dot{\phi} + \frac{\dot{\phi}^2}{\phi} &=& \frac{1}{(1 +
6\xi)} \frac{1}{\phi} \left[ 4 V(\phi) - \phi V^{\prime}(\phi) \right] ,
\label{Jfe1}
\eeqn
where overdots again denote time derivatives, and primes denote
$d/d\phi$.  As in~\cite{ngets}, we have assumed that the classical
background field $\phi$ is sufficiently homogenous, so that all spatial
derivatives become negligible. \\
\indent  Invoking the \lq\lq inflationary attractor"
assumption~\cite{psra}~\cite{SalBond} yields the approximate Jordan-frame
field equations~\cite{ngets}:
\beqn
\nonumber H^2 &\simeq& \frac{1}{3 \xi \phi^2} \> V(\phi) , \\
3H \dot{\phi} &\simeq& \frac{1}{(1 + 6 \xi)}\frac{1}{\phi} \left[ 4 V
(\phi) - \phi V^{\prime} (\phi) \right] .
\label{Jfe2}
\eeqn
Integration gives the closed-form solutions during the inflationary epoch:
\beqn
\nonumber \phi (t) &=& \phi_o \pm \sqrt{\frac{4 \lambda \xi}{3 (1 +
6\xi)^2}} \> v^2 \> t , \\
\frac{a(t)}{a_o} &=& \left( \frac{\phi (t)}{\phi_o} \right)^{(1 + 6\xi) /
4 \xi} \exp \left[ \frac{ (1 + 6 \xi)}{8 \xi v^2} \left( \phi_o^2 -
\phi^2 (t) \right) \right] ,
\label{aphi}
\eeqn
where $\phi_o$ and $a_o$ are values at the beginning of inflation.  In
the solution of $\phi (t)$, the $+$ corresponds to the \lq\lq new
inflation" initial conditions ($\phi_o \ll v$), and the $-$ corresponds
to \lq\lq chaotic inflation" initial conditions ($\phi_o \gg v$).  For
early times, then, under new inflation initial conditions the expansion
is predominantly quasi-power-law ($a(t) \propto t^{(1 + 6\xi) / 4\xi}$),
whereas under chaotic inflation initial conditions the expansion is
quasi-de Sitter ($a(t) \propto \exp ( \phi_o \sqrt{\lambda / 3\xi} \>
t)$).  As demonstrated in~\cite{ngets}, only the power-law expansion case
is affected by the discrepancy in $n_s$. \\
\indent  Working with the field equations in equation (\ref{Jfe2})
(appropriate for a first-order analysis)
and the inflationary solutions for $\phi (t)$ and $a (t)$ in equation
(\ref{aphi}), we
may evaluate the Jordan-frame slow-roll parameters ($\epsilon$, $\delta$,
and $\alpha$) for the new inflation scenario as:
\beqn
\nonumber  \epsilon &=& \frac{4 \xi}{1 + 6\xi} \>\> , \>\> \delta = 0 , \\
\alpha &=& \frac{2 \xi}{1 + 6\xi} \left( 1 + \frac{\epsilon}{\alpha}
\right) .
\label{aed}
\eeqn
The equation for $\alpha$ may be solved to give either $\alpha =
\epsilon$ or $\alpha = - \epsilon / 2$; yet for the new inflation initial
conditions, $\epsilon > 0$, and $\dot{\phi} > 0$, so only the solution
$\alpha = \epsilon$ may be chosen.  (Of course, we arrive at the same
result by considering that for IgI under either initial conditions,
$\alpha = \epsilon + \delta$ to first order, and, from equation
(\ref{aphi}), $\delta = 0$ for IgI to first order.)  From equation
(\ref{nuIgI}), this yields
\beq
\nu = \frac{1}{2} + \frac{2 \epsilon}{1 - \epsilon} =
\frac{1}{2} + \frac{8\xi}{1 + 2\xi} ,
\label{nuIgI2}
\eeq
or
\beq
n_{s} = 2 - 2 \nu = 1 - \frac{16\xi}{1 + 2\xi}
\label{nIgI2}
\eeq
for the spectral index from IgI, as calculated in the Jordan frame.  Note
that by using the new slow-roll expansion, based on the variables $u$ and
$z$ in equation (\ref{uvIgI}), this result for $n_s$ differs from that
calculated in the Jordan frame using the Einstein-frame formalism
of~\cite{SL}~\cite{LidLyth}.  (Compare equation (\ref{nIgI2})
with equation (47) in~\cite{ngets} or equation (27) in~\cite{DKplb}.) \\
\indent  To demonstrate the frame-independence of this result for $n_s$,
we may compare equation (\ref{nIgI2}) with a calculation in the Einstein
frame.  The action in equation (\ref{Sigi}) may be written in the
Einstein frame if we make the following conformal transformation (see,
{\it e.g.},~\cite{ngets}~\cite{HirMaeda}):
\beq
\tilde{g}_{\mu\nu} = \Omega^2 (x) g_{\mu\nu} \>\> , \>\> \Omega^2 (x) =
\kappa^2 \xi \phi^2 ,
\label{conftrans}
\eeq
where quantities in the Einstein frame are marked by a tilde.  As above
in section 2,
$\kappa^2 \equiv 8 \pi M_{pl}^{-2}$.  From the form of
the potential, $V (\phi)$, in equation (\ref{Sigi}), we may set
$\kappa^2 = (\xi v^2 )^{-1}$ for IgI.  If we
further define a new scalar field $\varphi$ and its potential $U$ by
\beq
\frac{d \varphi}{d \phi} \equiv \sqrt{\frac{1 + 6 \xi}{\kappa^2 \xi\phi^2 }}
\>\> , \>\> U \equiv \frac{1}{(\kappa^2 \xi \phi^2)^2} \> V
(\phi) , \eeq
then the action in this frame becomes
\beq
S = \int d^4 \tilde{x} \sqrt{- \tilde{g}} \left[ \frac{1}{2\kappa^2}
\tilde{R} - \frac{1}{2} \varphi_{; \> \mu} \varphi^{; \> \mu} -
U(\varphi) \right] ,
\eeq
giving the familiar field equations:
\beqn
\nonumber \tilde{H}^2 &=& \frac{\kappa^2}{3} \left[ \frac{1}{2} \left(
\frac{d \varphi}{d \tilde{t}} \right)^2 + U (\varphi) \right] , \\
\frac{d^2 \varphi}{d \tilde{t}^2} &+& 3 \tilde{H} \frac{d \varphi}{d
\tilde{t}} + \frac{d U}{d \varphi} = 0 ,
\label{Efe}
\eeqn
where $d \tilde{t} = \Omega (x) dt$, $d\tilde{\vec{x}} = d \vec{x}$,
$\tilde{a} (\tilde{t}) = \Omega (x) a (t)$, and $\tilde{H} =
\tilde{a}^{-1} d \tilde{a} / d \tilde{t}$. \\
\indent  Under new inflation conditions, $\Omega \propto \phi \propto t$;
given $a(t) \propto t^p$, with $p = (1 + 6 \xi) / 4 \xi$, the transformed
scale factor thus becomes $\tilde{a} (\tilde{t}) \propto
\tilde{t}^{\tilde{p}}$, with $\tilde{p} = (p + 1 ) / 2 = (1 + 10\xi) /
8\xi$.  From equations (\ref{epsdel}) and (\ref{Efe}), the slow-roll
parameters $\tilde{\epsilon}$ and $\tilde{\delta}$ may be evaluated as:
\beq
\tilde{\epsilon} = - \tilde{\delta} = \frac{1}{\tilde{p}} = \frac{8
\xi}{1 + 10\xi} .
\label{teps}
\eeq
Having made the conformal transformation of equation (\ref{conftrans}),
our theory is now in the form of an Einstein-Hilbert gravitational action
with a minimally-coupled scalar field; the appropriate $\nu$ from section
2 is therefore that given in equation (\ref{nuMSF}), which becomes
\beq
\tilde{\nu} = \frac{1}{2} + \frac{\tilde{\epsilon}}{1 - \tilde{\epsilon}} =
\frac{1}{2} + \frac{8\xi}{1 + 2\xi} ,
\label{tnu}
\eeq
giving, from equation (\ref{ns2}),
\beq
\tilde{n}_s = 1 - \frac{16\xi}{1 + 2\xi} .
\label{tn}
\eeq
Comparing equations (\ref{nIgI2}) and (\ref{tn}) it is clear that $n_s =
\tilde{n}_s$; the spectral index as calculated in the Jordan frame
matches the spectral index as calculated in the Einstein frame.

\section{Conclusion}
\indent  By developing an expansion in slow-roll parameters appropriate
to the complicated equation of motion for the gauge-invariant potential
$\Phi$, we have extended the usual Einstein-frame formalism for
calculating the spectral index into a form which may be used for GET
models of inflation.  The analysis has been conducted to first order only
in the slow-roll parameters.  It could be continued to second order by
following Stewart and Lyth's original derivation~\cite{SL}, that is, by
treating the spectrum $P_{\Phi}$ as adiabatic in the slowly-varying
$\epsilon$, $\delta$, $\alpha$, $\beta$, $\gamma$, and $\zeta$.  Yet,
because of the more complicated field equations for $\phi (t)$ and $a
(t)$ in the Jordan frame, as compared with the corresponding field
equations in the Einstein frame (compare, {\it e.g.}, equation
(\ref{Jfe1}) with (\ref{Efe})), an expansion to second order in the
Jordan-frame parameters would be exceedingly difficult to evaluate.
Instead, one should exploit the frame-independent nature of this
formalism, and evaluate the spectral index for such GETs in the Einstein
frame, as done in~\cite{ngets}. \\
\indent  The import of this work has been to remove the ambiguity, as
discussed in~\cite{ngets}, which
formerly plagued the evaluation of spectral indices for GET models of
inflation.  By seizing upon a generalization of the usual Einstein-frame
slow-roll expansion for $n_s$, which reduces to this
expansion following a conformal transformation, we have developed a
self-consistent formalism with which to calculate $n_s$ for GETs.

\section{Acknowledgments}
\indent  This research was conducted at Dartmouth College.  It is a
pleasure to thank Joseph Harris, Marcelo Gleiser, and John Walsh for their
hospitality.  This work was supported by an NSF Fellowship for
Pre-Doctoral Fellows.

%

%
%
\end{document}